\documentclass[journal]{IEEEtran}

\usepackage{xcolor}
\usepackage[colorlinks]{hyperref}
\usepackage[nolist]{acronym}
\usepackage[normalem]{ulem}
\usepackage[intlimits]{amsmath}
\usepackage{amsfonts}
\usepackage{amssymb}
\usepackage{booktabs}
\usepackage{gensymb}
\usepackage{tikz}
\usepackage{animate}
\usepackage{array}	
\usetikzlibrary{arrows}

\ifCLASSOPTIONcompsoc
\usepackage[caption=false,font=normalsize,labelfont=sf,textfont=sf]{subfig}
\else
\usepackage[caption=false,font=footnotesize]{subfig}
\fi

\hypersetup{
	colorlinks=true, % false: boxed links; true: colored links
	linkcolor=blue!70!black, % color of internal links
	citecolor=red!70!black, % color of links to bibliography
	filecolor=blue!70!black, % color of file links
	urlcolor=blue!70!black, % color of external links
}

\definecolor{PairedA}{RGB}{166, 206, 227}
\definecolor{PairedB}{RGB}{ 31, 120, 180}
\definecolor{PairedC}{RGB}{178, 223, 138} 
\definecolor{PairedD}{RGB}{ 41, 128,  35}   % made darker by 20%
\definecolor{PairedE}{RGB}{251, 154, 153}
\definecolor{PairedF}{RGB}{182,  21,  22}   % made darker by 20%
\definecolor{PairedG}{RGB}{253, 191, 111}
\definecolor{PairedH}{RGB}{255, 127,   0}
\definecolor{PairedI}{RGB}{202, 178, 214}
\definecolor{PairedJ}{RGB}{106,  61, 154}
\definecolor{PairedK}{RGB}{255, 255, 153}
\definecolor{PairedL}{RGB}{177,  89,  40}

\providecommand{\J}{\ensuremath{\mathrm{j}}}        % imaginary unit
\providecommand{\RE}{\ensuremath{\mathrm{Re}}}  % real part of...
\providecommand{\IM}{\ensuremath{\mathrm{Im}}}  % imaginary part of...

\providecommand{\Quot}[1]{``{#1}"}              % quotation
\providecommand{\V}[1]{\boldsymbol{#1}}         % vector
\providecommand{\M}[1]{\mathbf{#1}}             % matrix
\providecommand{\T}[1]{\mathrm{#1}}             % upright text (subindex, superindex,..)
\providecommand{\UV}[1]{\V{\hat{#1}}}           % unit vector
\providecommand{\OP}[1]{{\mathcal{#1}}}         % operator
\providecommand{\contragradient}[1]{\widetilde{#1}} % contragradient

\providecommand{\trans}{\mathrm{T}}

\providecommand{\srcRegion}{\ensuremath{\varOmega}} 
\providecommand{\basisFcn}{\V{\psi}}

\providecommand{\Ivec}{\ensuremath{\M{I}}}

\providecommand{\Jr}{\ensuremath{\V{J}\left(\V{r}\right)}}
\providecommand{\CR}{\ensuremath{\M{C}\left(R\right)}}
\providecommand{\CRG}{\ensuremath{\M{C}\left(R \in G\right)}}
\providecommand{\DR}{\ensuremath{\M{D}\left(R\right)}}
\providecommand{\DRG}{\ensuremath{\M{D}\left(R \in G\right)}}
\providecommand{\Imat}{\ensuremath{\left[\M{I}\right]}}

\newcommand{\ie}{\textit{i.e.}{}}
\newcommand{\eg}{\textit{e.g.}{}}

\newcommand\figwidth{8.9} % in cm
\newcommand\subfigwidth{4.3} % in cm
\usepackage{graphicx}

\newacro{MoM}[MoM]{method of moments}
\newacro{MOO}[MOO]{multiobjective optimization}
\newacro{CM}[CM]{characteristic mode}
\newacro{PEC}[PEC]{perfect electric conductor}
\newacro{PMC}[PMC]{perfect magnetic conductor}
\newacro{EP}[EP]{eigenvalue problem}
\newacro{GEP}[GEP]{generalized eigenvalue problem}
\newacro{EFIE}[EFIE]{electric field integral equation}
\newacro{SVD}[SVD]{singular value decomposition}
\newacro{RWG}[RWG]{Rao-Wilton-Glisson}
\newacro{EM}[EM]{electromagnetic}
\newacro{DOF}[DOF]{\mbox{degrees-of-freedom}}

\usepackage[textsize=tiny,backgroundcolor=red,linecolor=red,textwidth=2cm]{todonotes}

\newif\ifarxiv
\arxivfalse % TAP version
\begin{document}
% ========================================================================
% PAPER
% ========================================================================

\title{Modal Tracking Based on Group Theory}
\author{Michal~Masek, Miloslav~Capek, \IEEEmembership{Senior Member, IEEE}, Lukas~Jelinek, and Kurt~Schab, \IEEEmembership{Member, IEEE}
\thanks{Manuscript received  December 10, 2018; revised May 13, 2019. This work has been supported by the Grant Agency of the Czech Technical University in Prague under grant~\mbox{SGS19/168/OHK3/3T/13}. The work of M.~Capek was supported by the Ministry of Education, Youth and Sports through the project CZ.02.2.69/0.0/0.0/16\_027/0008465.}% <-this % stops a space
\thanks{M. Masek, M. Capek, and L. Jelinek are with the Czech Technical University in Prague, Prague, Czech Republic (e-mails: \{michal.masek; miloslav.capek; lukas.jelinek\}@fel.cvut.cz).}% <-this % stops a space
\thanks{K. Schab is with the Department of Electrical Engineering, Santa Clara University, Santa Clara, USA (e-mail: kschab@scu.edu).}% <-this % stops a space
}    
%\markboth{Journal of \LaTeX\ Class Files,~Vol.~XX, No.~XX, \today}{Masek \MakeLowercase{\textit{et al.}}: Modal Tracking Based on Group Theory}

\maketitle

\begin{abstract}
Issues in modal tracking in the presence of crossings and crossing avoidances between eigenvalue traces are solved via the theory of point groups. The von~Neumann-Wigner theorem is used as a key factor in predictively determining mode behavior over arbitrary frequency ranges. The implementation and capabilities of the proposed procedure are demonstrated using characteristic mode decomposition as a motivating example. The procedure is, nevertheless, general and can be applied to an arbitrarily parametrized eigenvalue problems. A treatment of modal degeneracies is included and several examples are presented to illustrate modal tracking improvements and the immediate consequences of improper modal tracking. An approach leveraging a symmetry-adapted basis to accelerate computation is also discussed. A relationship between geometrical and physical symmetries is demonstrated on a practical example.
\end{abstract}


%\textbf{\small{\emph{Index Terms} --- Eigenvalues and eigenfunctions, antenna theory, numerical analysis, electromagnetic modeling, coupled mode analysis.}}
%\IEEEpeerreviewmaketitle

%xxxxxxxxxxxxxxxxxxxxxxxxxxxxxxxxxxxxxxxxxxxxxxxxxxxxxxxxxxxxxxxxxxxxxx
\section{Introduction}
\label{sec:intro}

\IEEEPARstart{M}{odal} tracking is an important part of every procedure dealing with parametrized eigenvalue problems. In antenna theory, eigenvalue problems -- for example, the problem defining \acp{CM}~\cite{HarringtonMautz_TheoryOfCharacteristicModesForConductingBodies} -- are commonly parametrized in frequency and solved at a finite set of discrete frequency points. Therefore, tracking is required to order and associate modes across the frequency band of interest so as to obtain modal quantities which are smooth functions of frequency. A common way of dealing with modal tracking is to employ correlation between modes \cite{CapekHazdraHamouzEichler_AMethodForTrackingCharNumbersAndVectors,RainesRojas_WidebandCharacteristicModeTracking, LudickJakobusVogel_AtrackingAlgorithmForTheEigenvectorsCalculatedWithCM, MiersLau_WideBandCMtrackingUtilizingFarFieldPatterns, SafinManteuffel_AdvancedEigenvalueTrackingofCM}. Despite the success of correlation-based tracking~\cite{CapekEtAl_ValidatingCMsolvers}, there are, however, scenarios in which this technique fails to provide reliable results. One specific example where correlation-based tracking becomes difficult in the vicinity of accidental degeneracies or crossing avoidances~\cite{vonNeumannWigner_OnTheBehaviourOfEigenvaluesENG,SchabEtAl_EigenvalueCrossingAvoidanceInCM}. Point group theory~\cite{McWeeny_GroupTheory,Cornwell_Group_Theory_Intro,Cornwell_Group_Theory_1_2} has recently been used to provide ground truth for determining whether modal degeneracies may exist by means of the von Neumann-Wigner theorem and to show that if the object under study exhibits no symmetry, traces of the characteristic numbers, as functions of frequency, cannot cross~\cite{SchabBernhard_GroupTheoryForCMA,vonNeumannWigner_OnTheBehaviourOfEigenvaluesENG}. Simultaneously, it has been shown that when symmetries are present, modes can be divided into separate unique sets called irreducible representations (irreps) and that within these sets, modal crossings are limited by simple, known features of each irrep. The problem of how to detect symmetries and assign modes to irreps in the general case of multidimensional irreps encountered in bodies with non-abelian symmetry groups~\cite{McWeeny_GroupTheory} has not been solved.  Our goal in this work is to establish a robust modal tracking implementation based on the fundamental modal crossing rules discussed in \cite{SchabBernhard_GroupTheoryForCMA} capable of working with non-symmetric, abelian symmetric, and non-abelian symmetric structures.

To accomplish this, a procedure for performing classification of modes is developed for surfaces having any known symmetry point group. An application for broadband characteristic mode tracking, which includes two approaches, is demonstrated. The first, an \textit{a posteriori} approach, deals with previously calculated modes and their assignment into irreps. The second, an \textit{a priori} approach, uses a symmetry-adapted basis to block-diagonalize the underlying operator and divide the problem into a series of smaller parts which are solved independently. Each part spans a separate eigenmode subspace corresponding to a given irrep~\cite{Knorr_1973_TCM_symmetry}. Computation speed is remarkably accelerated in this latter approach due to the cubic dependence of computation time on the number of discretization elements. Despite the aforementioned differences, both methods divide modes into sets where traces of the eigenvalues cannot cross, automatically solving the eigenvalue crossing/crossing avoidance issue. A framework based on group theory presented in this paper is numerically implemented on \ac{RWG} basis functions~\cite{RaoWiltonGlisson_ElectromagneticScatteringBySurfacesOfArbitraryShape} and its capabilities are demonstrated on characteristic mode decomposition~\cite{HarringtonMautz_TheoryOfCharacteristicModesForConductingBodies}. It should, however, be noted that the results are applicable to other modal decompositions or alternative choices of basis functions as well.

The paper is organized as follows. Background theory is reviewed in Section~\ref{sec:background} where characteristic modes are introduced as an example generalized eigenvalue problem in Section~\ref{sec:background:charModes}. Point group theory and modal classification into irreps is described in Section~\ref{sec:background:pointGroup} and the utilization of a symmetry-adapted basis to reduce the original problem is shown in Section~\ref{sec:background:symAdaptBasis}. Section~\ref{sec:implementation} describes a practical implementation and basic results using an illustrative example. Further examples of results are presented in Section~\ref{sec:examples}. Section~\ref{sec:discussion} is dedicated to a discussion of several related topics: the importance of modal tracking in Section~\ref{sec:discusion:importance}, the effective generation of a symmetry-adapted basis in Section~\ref{sec:discussion:symAdaptBasis}, the reduction of computational time of modal decomposition in Section~\ref{sec:discussion:speedup} and the relationship between geometrical and physical symmetries in Section~\ref{sec:discussion:PECPMC}. Section~\ref{sec:conclusion} concludes the paper.

%xxxxxxxxxxxxxxxxxxxxxxxxxxxxxxxxxxxxxxxxxxxxxxxxxxxxxxxxxxxxxxxxxxxxxx
\section{Background Theory}
\label{sec:background}

The following sections briefly summarize the theory of characteristic modes (Section~\ref{sec:background:charModes}) and point group theory (Sections~\ref{sec:background:pointGroup},~\ref{sec:background:symAdaptBasis}). Details related to modal tracking are emphasized.

%xxxxxxxxxxxxxxxxxxxxxxxxxxxxxxxxxxxxxxxxxxxxxxxxxxxxxxxxxxxxxxxxxxxxxx
\subsection{Characteristic Modes}
\label{sec:background:charModes}

Characteristic modes~(CMs)~\cite{HarringtonMautz_TheoryOfCharacteristicModesForConductingBodies} form a set of orthogonal solutions to a \ac{GEP}~\cite{Wilkinson_AlgebraicEigenvalueProblem}
\begin{equation}
	\label{eq:CMOP}
	\OP{X} \V{J}_n \left(\V{r}\right) = \lambda_n \OP{R} \V{J}_n \left(\V{r}\right),
\end{equation}
in which~$\OP{Z} = \OP{R} + \J \OP{X}$ is the impedance operator~\cite{HarringtonMautz_TheoryOfCharacteristicModesForConductingBodies} defined for a perfectly electrically conducting body $\srcRegion$ as
\begin{equation}
\label{eq:EFIE}
	\OP{Z} \V{J} \left(\V{r}\right) = \UV{n} \times \UV{n} \times \V{E} \left( \V{J} \left(\V{r}\right) \right).
\end{equation}
In~\eqref{eq:EFIE}, $\V{r} \in \srcRegion$, $\UV{n}$~denotes a unit normal to surface~$\srcRegion$ and $\V{E}$ denotes the scattered electric field produced by the electric current $\V{J} \left(\V{r}\right)$~\cite{Harrington_TimeHarmonicElmagField}. Each characteristic vector~$\V{J}_n \left(\V{r}\right)$ and its corresponding characteristic number~$\lambda_n$ define a unique eigensolution, completely independent of excitation, as being a sole function of geometry~$\srcRegion$ and angular frequency~$\omega$. The most appealing property of \acp{CM} is their ability to diagonalize the impedance operator~\eqref{eq:EFIE} as

\begin{equation}
	\label{eq:Zdiag}
	\dfrac{\langle \V{J}_m \left(\V{r}\right), \OP{Z} \V{J}_n \left(\V{r}\right) \rangle}{ \langle \V{J}_n \left(\V{r}\right), \OP{R} \V{J}_n \left(\V{r}\right) \rangle} =  \left( 1 + \J \lambda_n \right) \delta_{mn},
\end{equation}
where $\langle \cdot, \cdot \rangle$ is the inner product
\begin{equation}
    \langle \V{A} \left(\V{r}\right), \V{B} \left(\V{r}\right) \rangle = \int_\varOmega \V{A}^*(\V{r})\cdot\V{B}(\V{r})\mathrm{d}\V{r}
\end{equation}
 and $\delta_{mn}$ is the Kronecker delta.

The solution to~\eqref{eq:CMOP} is typically approached using the \ac{MoM}~\cite{Harrington_FieldComputationByMoM} which recasts the original integro-differential problem into a matrix problem, \ie{},
\begin{equation}
  	\label{eq:CMmat}
	\M{X} \Ivec_n = \lambda_n \M{R} \Ivec_n,
\end{equation}
using a set of $M$ basis functions~$\left\lbrace\basisFcn_m \left( \V{r} \right)\right\rbrace$ to approximate the current density as
\begin{equation}
	\label{eq:Jexpansion}
	\Jr \approx \sum_{m=1}^M I_m \basisFcn_m (\V{r})
\end{equation}
and performing Galerkin testing~\cite{PetersonRayMittra_ComputationalMethodsForElectromagnetics}, \ie{},
\begin{equation}
	\label{eq:Zmatrix}
	\M{R} + \J \M{X} = \left[ z_{mn} \right] = \left[ \langle \basisFcn_m \left( \V{r} \right) , \OP{Z} \basisFcn_n \left( \V{r} \right) \rangle \right].
\end{equation}
Within the discretized (matrix) form, the unknowns are column vectors of expansion coefficients~\mbox{$\Ivec \in \mathbb{C}^{M\times 1}$}, see~\cite{Harrington_FieldComputationByMoM}. If~$N$~characteristic modes are calculated, they form a matrix
~\mbox{$\Imat = [\Ivec_1, \dots, \Ivec_n, \dots, \Ivec_N] \in \mathbb{C}^{M\times N}$} with an associated vector of eigenvalues~\mbox{$\V{\lambda} = \left[ \lambda_1, \dots, \lambda_n,\dots, \lambda_N \right]$}. Notice that, theoretically,~$\Imat$ is a square matrix (\mbox{$M = N$}), however, in practice, the number of modes is typically limited to~\mbox{$N \ll M$} due to numerical constraints \cite{CapekEtAl_ValidatingCMsolvers}.

The necessity of modal tracking arises from the fact that the \ac{GEP}~\eqref{eq:CMOP} is parametrized by angular frequency~$\omega$. Therefore, if~\eqref{eq:CMmat} is to be solved at $F$ unique frequency points, its solution must be evaluated at each frequency point independently. If collected together, the resulting matrix structure of the eigenvalues reads~\mbox{$\left[ \V{\lambda} \right] \in \mathbb{R}^{F \times N}$}. Generally, there is no guarantee that all of the eigenvalues (and similarly for eigenvectors) in one column of~$\left[ \V{\lambda} \right]$ will belong to the same mode, \ie{}, the modal data are not properly tracked over frequency. Consequently, the frequency behavior of the individual characteristic modes cannot be effectively studied. 

\begin{figure}
	\includegraphics[width=\figwidth cm]{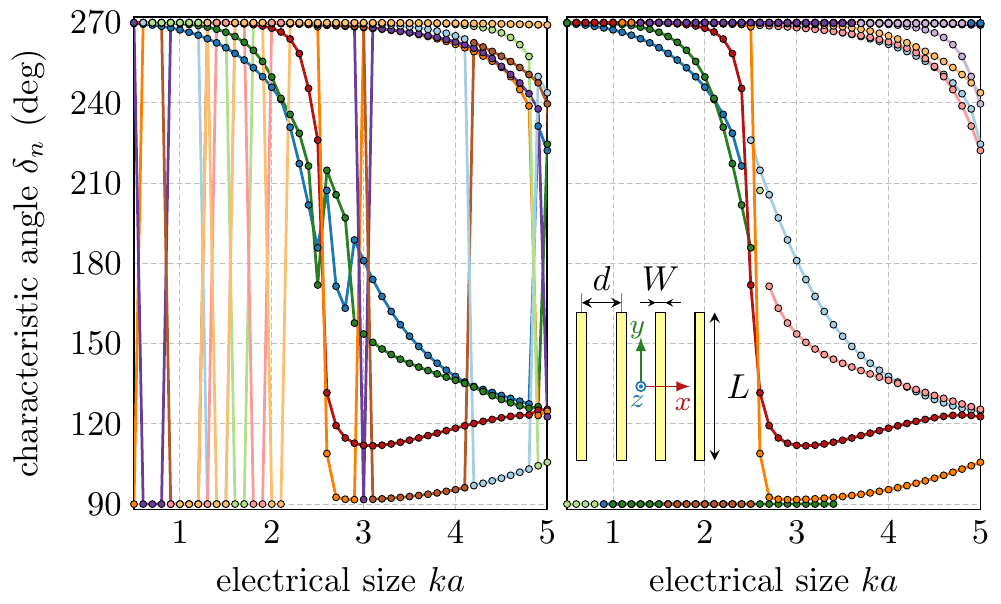}
	\caption{Raw untracked (left) and correlation tracked (right) traces of characteristic angles. Point markers indicate computed frequency samples.}
	\label{fig:trackingDemonstration}
\end{figure} 

The effects of insufficient tracking are demonstrated in Fig.~\ref{fig:trackingDemonstration} on a practical example of four parallel dipoles made of a \ac{PEC} with length~$L$, width~\mbox{$W = L/25$}, and separation distance~\mbox{$d=L/2$}. Frequency is scaled as electrical size $ka$ ($k$ is the free space wave-number and $a$ is the radius of the smallest circumscribing sphere) and spans from \mbox{$ka=0.5$} to \mbox{$ka = 5$} with \mbox{$F = 46$} equidistantly distributed frequency samples. The first \mbox{$N=15$} characteristic modes were calculated in AToM~\cite{atom} using \mbox{$M=496$} \ac{RWG} basis functions~\cite{RaoWiltonGlisson_ElectromagneticScatteringBySurfacesOfArbitraryShape}. The left panel of Fig.~\ref{fig:trackingDemonstration} shows untracked eigenvalue modal traces, while the right panel of the same figure shows modal eigenvalue traces tracked using correlation between two neighboring frequencies~\cite{CapekHazdraHamouzEichler_AMethodForTrackingCharNumbersAndVectors, LudickJakobusVogel_AtrackingAlgorithmForTheEigenvectorsCalculatedWithCM}. The characteristic numbers~$\lambda_n$ are rescaled as characteristic angles~$\delta_n$,~\cite{Garbacz_TCMdissertation, Newman_SmallAntennaLocationSynthesisUsingCharacteristicModes}
\begin{equation}
	\label{eq:charAngles}
	\delta_n = 180 \degree \left( 1 - \frac{1}{\pi} \arctan \left(\lambda_n \right) \right).
\end{equation}
The results presented in Fig.~\ref{fig:trackingDemonstration} demonstrate the impact of modal tracking as well as some of the shortcomings of current and field correlation based tracking algorithms previously discussed in the literature~\cite{RainesRojas_WidebandCharacteristicModeTracking, MiersLau_WideBandCMtrackingUtilizingFarFieldPatterns, SafinManteuffel_AdvancedEigenvalueTrackingofCM}. Specifically, it is observed that tracking problems commonly occur in regions where modal eigenvalue traces cross or come very near one another. Some traces are also unpredictably stopped while new traces arise when the correlation value between modes at adjacent frequency samples do not exceed some user-defined threshold.

While the preceding discussion focused on characteristic modes, many problems in computational electromagnetics rely on the broadband solution of parametrized eigenvalue problems in the form of
\begin{equation}
	\label{eq:arbGEP}
	\M{A} \left(p \right) \Ivec_n = \lambda_n \M{B} \left(p\right) \Ivec_n,
\end{equation}
for which the same consideration of broadband tracking must be given for the proper and consistent interpretation of results.

%----------------------------------------------------------------------
\subsection{Point Group Theory and Classification of Modes}
\label{sec:background:pointGroup}

This section introduces necessary elements of the theory of symmetry point groups~\cite{McWeeny_GroupTheory,Cornwell_Group_Theory_Intro,Cornwell_Group_Theory_1_2} as well as key results from group theory which are relevant for the symmetry classification of eigenmodes and eigenvalue tracking.

To begin, assume a surface~$\varOmega$ which supports a surface current density $\Jr$. The symmetry point group $G$ of this object is the set \mbox{$G = \left\lbrace R_i\right\rbrace$} of all symmetry operations~$R$ (\eg{}, identity~$\T{E}$, $n$-fold rotations~$\T{C}_n$, reflections~$\T{\sigma}$, and their products) to which the object is invariant. Let the vector space~$V$ contain all possible complex current distributions~$\Jr$ on this object. The operation~$R$ maps~$V$ onto itself, \ie{},
\begin{equation}
	\label{eq:Rmapping}
	R\Jr = \M{T}\left(R\right) \V{J} \left( \M{T}^{-1}\left(R\right) \V{r} \right) \in V, \quad \forall \Jr \in V,
\end{equation}
or, equivalently, the operation~$R$ transforms one current on~$\varOmega$ into another current on~$\varOmega$. The matrix~$\M{T}\left(R\right)$ in~\eqref{eq:Rmapping} is a \mbox{$3 \times 3$} coordinate transformation matrix corresponding\footnote{We assume that $\Jr$ is a polar vector.} to the operation~$R$. If a finite basis~$\{\basisFcn_m (\V{r})\}$ is chosen to approximate the space~$V$ according to~\eqref{eq:Jexpansion}, and if the basis is chosen to preserve the symmetry of~$\varOmega$ (\ie{}, the object~$\varOmega$ is represented by a mesh which preserves all symmetries of the object), then by~\eqref{eq:Rmapping},
\begin{equation}
	\label{eq:Cmapping:sum}
	R\basisFcn_n\left(\V{r}\right) = \sum_{m=1}^{M} c_{mn}(R) \basisFcn_m\left(\V{r}\right).
\end{equation}
Note here that $R$ is any element of group $G$, and that the coefficients $c_{mn}(R)$ can be collected in a mapping matrix \mbox{$\CR = \left[ c_{mn} (R) \right]$}.  Thus, by~\eqref{eq:Jexpansion}, the effect of $R$ on the current $\Jr$ is realized by
\begin{equation}
	\label{eq:Cmapping}
	R\Ivec = \CR \Ivec,
\end{equation}
where~$R \Ivec$ is the vector of expansion coefficients corresponding to~$R\Jr$. The set of mapping matrices~$\left\lbrace\CRG\right\rbrace$ define a matrix representation of~$G$. The structure and nature of these matrices is dependent of the basis chosen. Consider~$N$ current expansion vectors $\Ivec_n$ which form a modal basis. These can, \eg{}, be \acp{CM} defined by~\eqref{eq:CMmat}.  Equivalently, as shown in~\eqref{eq:Cmapping:sum}, the effect of $R$ on a particular mode can be expressed as
\begin{equation}
	\label{eq:Dmapping:sum}
  	R \Ivec_n = \sum_{m=1}^N d_{mn}(R) \Ivec_m,
\end{equation}
where the matrix~\mbox{$\DR = \left[ d_{mn}(R) \right]$} is a matrix representation of $R$ in the modal basis. Arranging all modal vectors~$\Ivec$ and $R \Ivec$ into matrices~$\Imat$ and~$\CR\Imat$, relation~\eqref{eq:Dmapping:sum} can also be written as
\begin{equation}
  \label{eq:Dmapping:matrices}
  \Imat \DR = \CR \Imat.
\end{equation}

For a symmetrical object, subsets of modes exclusively map onto themselves with each operation $R \in G$,~\cite{McWeeny_GroupTheory}. This is observed, after the specific ordering of columns in the matrix~$\Imat$, as a block diagonalization of $\DR$, 
\begin{equation}
	\label{eq:Dmatrix}
    \left[
    \begin{array}{*{20}{c}}
    \M{D}_1(R)&0&0&0\\
    0&\M{D}_2(R)&0&0\\
    0&0& \ddots &0\\
    0&0&0&\M{D}_Q(R)
    \end{array}\right] = {\Imat}^{-1} \CR \Imat.
\end{equation}
Each collection of block matrices~\mbox{$\left\{\M{D}_q \left( R \in G \right)\right\}$}, for a certain block number~\mbox{$q \in \left\{ 1, \dots, Q\right\}$}, and its corresponding subset of the modal basis, is known as an irreducible representation (irrep) of the group~$G$,~\cite{McWeeny_GroupTheory}. The characters \mbox{$\chi_q\left(R\right) = \T{tr}\left(\M{D}_q\left(R\right)\right)$} are commonly used to fully characterize the group~$G$ and to classify modes within irreps. Note that the block matrices of a particular irrep may be repeated multiple times along the diagonal of~$\DR$, which means that separate sets of modes may belong to the same irrep but map only onto themselves. The dimension of the block matrices in an irrep is known as the dimension~$g^\alpha$ of that irrep, where the superindex~$\alpha$ is used throughout the paper to denote different irreps\footnote{Standard designations for irreps are used in this paper: $\T{A}$ and $\T{B}$ for one-dimensional irreps, $\T{E}$ for two-dimensional irreps and $\T{T}$ for three-dimensional irreps, see, \eg{},~\cite{McWeeny_GroupTheory}.}. For the rest of the paper, a particular block, \eg{}, $\M{D}_q\left(R\right)$, corresponding to an irrep $\alpha$ is denoted $\M{D}^{\alpha}\left(R \right)$.

Identifying irreps, the von Neumann-Wigner theorem~\cite{vonNeumannWigner_OnTheBehaviourOfEigenvaluesENG} states that modal degeneracies can only occur between modes of the same irrep up to the dimension of that irrep. Given that the difficulties in tracking characteristic modes over frequency largely stem from the accurate identification of degeneracies~\cite{SchabEtAl_EigenvalueCrossingAvoidanceInCM} (eigenvalue crossings), this result is of great utility and provides an analytic ground truth for verifying mode tracking algorithms. For example, this result states that two eigenvalue traces may not cross if both belong to modes within the same Abelian (one-dimensional) irrep.  If a modal tracking algorithm under test outputs eigenvalue traces that do cross in spite of this, we can deduce that the modal tracking was not performed correctly.  A similar situation occurs in higher-dimension irreps with continuously degenerated modes, as discussed in Section~\ref{sec:examples:equilateralTriangle}.

%----------------------------------------------------------------------
\subsection{Symmetry-Adapted Basis and Reduced Problems}
\label{sec:background:symAdaptBasis}

The presence of symmetries is not limited to the \textit{a posteriori} classification of modes as described above, but it can be used to produce \Quot{symmetry-adapted} eigenvalue problems that directly produce modes corresponding to a given irrep. To that point, imagine that matrices~\mbox{$\M{D}^\alpha \left( R\right)$} are known. Then, according to~\cite{McWeeny_GroupTheory}, the left hand side of
\begin{equation}
    \label{eq:symmetryAdaptedVectors}
  	\V{V}_{i}^\alpha = \frac{g^\alpha}{g} \sum_{R \in G} \contragradient{d}^\alpha_{ii} \left(R \right) \M{C} \left(R \right) \V{v},
\end{equation}
is a transformed version of the arbitrary vector $\V{v}$ which is \Quot{symmetry-adapted} to the $i$-th degeneracy (dimension) of irrep~$\alpha$. Here,~\mbox{$\contragradient{\M{D}} = \left(\M{D}^{-1}\right)^\T{T}$} stands for a contra-gradient representation, and \mbox{$g = \sum_\alpha \left(g^\alpha\right)^2$ is the order of group~$G$}. Here we assign~$\V{v}$ as the columns of an $M \times M$ identity matrix, so that
\begin{equation}
  	\label{eq:symmetryAdaptedBasis}
  	\V{\rho}_{i}^\alpha = \frac{g^\alpha}{g} \sum_{R \in G} \contragradient{d}_{ii}^\alpha \left(R \right) \M{C} \left(R \right)
\end{equation}
produces matrices~$\V{\rho}_{i}^\alpha$ whose columns are \Quot{symmetry-adapted} vectors. Naturally, there are only \mbox{$\eta^\alpha = \T{rank}\left(\M{\Gamma}^\alpha_{i}\right) \leq M$} linearly independent columns in each matrix~$\V{\rho}_{i}^\alpha$. Removal of the linearly dependent columns from the matrices~$\V{\rho}_{i}^\alpha$ produces matrices~$\M{\Gamma}_i^\alpha$ of size~\mbox{$M \times \eta^\alpha$} which are able to modify matrix operators of the physical system at hand so that their eigenvectors belong solely\footnote{In the case of multidimensional representation, all matrices~\mbox{$\M{\Gamma}_i^\alpha$} for \mbox{$i \in \left\{ 1,\ldots, g^\alpha \right\}$} must be employed to obtain all modes belonging to an irrep~$\alpha$.} to an irrep~$\alpha$. Explicitly, having a generalized eigenvalue problem~\eqref{eq:arbGEP}, a solution subspace of eigenvectors belonging to an irrep~$\alpha$ is found, assuming that in this irrep, due to the presence of symmetry, an eigenvector~$\Ivec \in \mathbb{C}^{M\times 1}$ can be composed as $\M{\Gamma}_i^\alpha \widehat{\Ivec}_i^\alpha$, where~\mbox{$\widehat{\Ivec}_i^\alpha \in \mathbb{C}^{\eta^\alpha\times 1}$}. Substituting into~\eqref{eq:arbGEP} and multiplying from the left by~$\left(\M{\Gamma}_i^\alpha\right)^\T{T}$ results in
\begin{equation}
  	\label{eq:reduced:GEP}
  	\widehat{\M{A}}_i^\alpha \widehat{\Ivec}_{i,n}^\alpha = \lambda_{i,n}^\alpha \widehat{\M{B}}_i^\alpha \widehat{\Ivec}_{i,n}^\alpha,
\end{equation}
where
\begin{align}
	\widehat{\M{A}}_i^\alpha &= \left(\M{\Gamma}_i^\alpha\right)^\trans \M{A} \M{\Gamma}_i^\alpha, \\
    \widehat{\M{B}}_i^\alpha &= \left(\M{\Gamma}_i^\alpha\right)^\trans \M{B} \M{\Gamma}_i^\alpha.
\end{align}
The modes generated by the eigenvalue problem~\eqref{eq:reduced:GEP} belong solely to the irrep~$\alpha$.

The matrices~\mbox{$\M{D}^\alpha \left( R \in G \right)$} must be known in order to evaluate the reduction matrices~$\M{\Gamma}_i^\alpha$. It is important, however, to realize that these can be obtained from any eigenvalue decomposition using~\eqref{eq:Dmatrix} on any object with the same symmetries. This initial object can, therefore, be chosen to have a minimal number of basis functions, making the evaluation of the matrices~\mbox{$\M{D}^\alpha \left( R \in G \right)$} computationally inexpensive.

%xxxxxxxxxxxxxxxxxxxxxxxxxxxxxxxxxxxxxxxxxxxxxxxxxxxxxxxxxxxxxxxxxxxxxx
\section{Practical Evaluation of Required Matrices}
\label{sec:implementation}

An implementation of the methodology introduced in the previous sections is shown on a test case, depicted in Fig.~\ref{fig:simpleStructure}, consisting of a triangularized domain~$\srcRegion$, \cite{deLoeraRambauSantos_Triangulations}, for which \mbox{$N=5$} \ac{RWG} basis functions are generated. Construction of the matrices $\M{C}(R)$, $\DR$, and $\V{\Gamma}_i^\alpha$ is demonstrated using this simple object with symmetry group\footnote{It is supposed that the set of operations~$G$ is known for a given structure.}~\mbox{$G = \left\lbrace \T{E}, \sigma_\T{v}^{yz}, \sigma_\T{v}^{xz}, \T{C}_{2}^z\right\rbrace$}.

%----------------------------------------------------------------------
\subsection{Construction of Matrices~$\M{C}(R)$}
\label{sec:implementation:Cmat}

A necessary step to classify modes according to their symmetry properties is to construct mapping matrices~$\CRG$. Here we assume that each basis function is mapped onto exactly one basis function under each symmetry operation within the symmetry group of the system. Under this assumption, the behavior of symmetry operation~$R$ is tested individually by~\eqref{eq:Rmapping} for all doublets of basis functions as
\begin{equation}
	\label{eq:SymTest}
    \begin{split}
	\gamma_{m}\left(R\right) = n: \;  \basisFcn_n \left(\V{r}\right) = s_m\left(R\right)\,\M{T}\left(R\right) \basisFcn_m \left({\M{T}}^{-1}\left(R\right) \V{r}\right),
    \end{split}
\end{equation}
where the vector~\mbox{$\V{\gamma}\left(R\right)$} contains an integer~$n$ at each position~$m$ so that the basis functions~$\basisFcn_m$ and~$\basisFcn_n$ map onto one another by the operation~$\M{T}\left(R\right)$ with a sign~\mbox{$s_m = \pm 1$}. The vectors~$\V{\gamma}\left(R\right)$ and~$\V{s}\left(R\right)$ are constructed for all symmetry operations $R \in G$.

\begin{figure}
    \includegraphics[width=\figwidth cm]{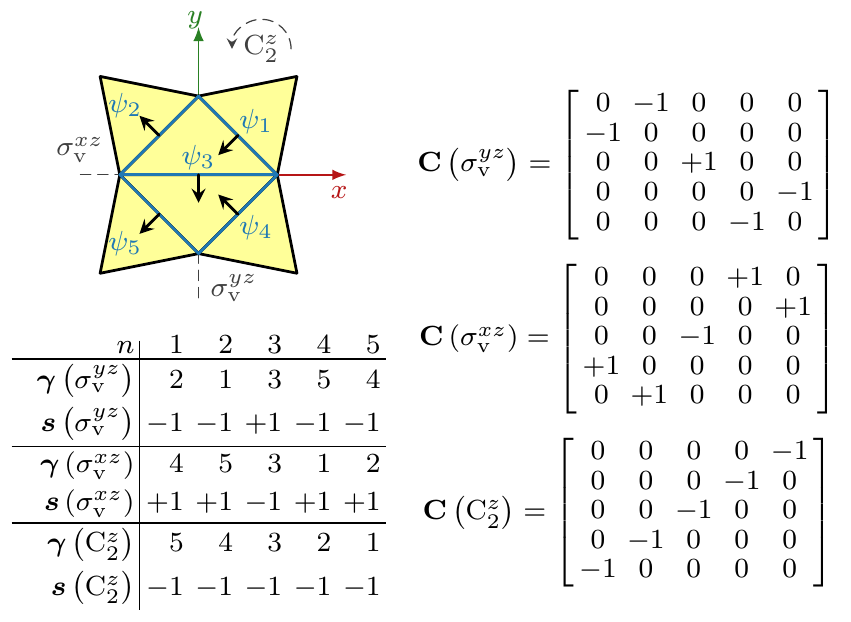}
    \caption{An illustration of a simple symmetric structure with depicted RWG basis functions~$\left\{\basisFcn_n \left(\V{r}\right)\right\}$. The structure belongs to point group $\T{C}_{2\T{v}}$~\cite{McWeeny_GroupTheory} with symmetry operations \mbox{$G = \left\lbrace \T{E}, \sigma_\T{v}^{yz}, \sigma_\T{v}^{xz}, \T{C}_{2}^z\right\rbrace$}. Vectors~$\V{\gamma}\left(R \in G\right)$, $\V{s}\left(R \in G\right)$, and matrices $\M{C}\left(R \in G\right)$ are shown. The matrix $\M{C}\left(\T{E}\right)$ is an identity matrix and is omitted.}
    \label{fig:simpleStructure}
\end{figure}

With all vectors~$\V{\gamma}\left(R \in G\right)$ and~$\V{s}\left(R \in G\right)$ known, the orthonormal mapping matrices $\CRG  = \left[ c_{mn} (R) \right]$, see~\eqref{eq:Cmapping:sum} and~\eqref{eq:Cmapping}, are constructed as
\begin{equation}
	\label{eq:construcCMatrix}
    c_{mn}\left(R\right) = \left\{ {\begin{array}{*{30}{l}}
			s_m\left(R\right) & \T{if}~n = \gamma_m\left(R\right), \\
			0 & \T{otherwise}.
			\end{array}} \right.
\end{equation}
A particular example of this procedure is shown in Fig.~\ref{fig:simpleStructure}, including the vectors~$\V{\gamma}\left(R\right)$ and $\V{s}\left(R\right)$, and the resulting matrices~$\CRG$. Note that the trivial results associated with the identity operator $\T{E}$ are omitted.

%----------------------------------------------------------------------
\subsection{Construction of Matrices~$\M{D}(R)$}
\label{sec:implementation:Dmat}

\begin{figure}
	\includegraphics[width=\figwidth cm]{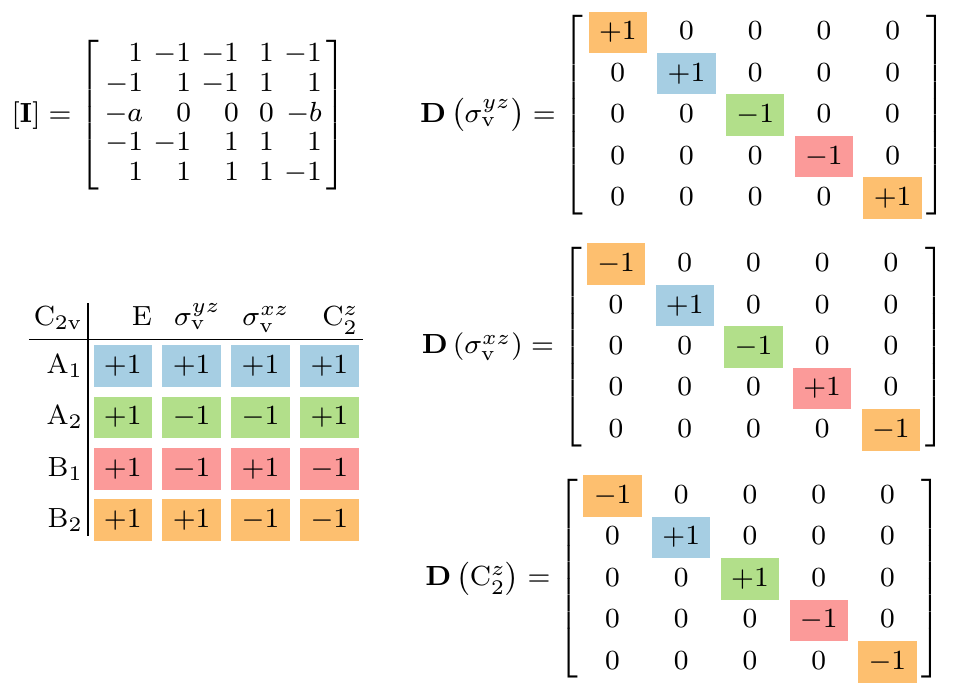}
	\caption{Eigenvectors~$\Imat$ computed by~\eqref{eq:CMmat} at \mbox{$ka = 0.5$} and corresponding matrices $\DRG$ are shown for the structure depicted in Fig.~\ref{fig:simpleStructure}. Each vector is normalized resulting in \mbox{$a = 1.764$} and \mbox{$b = 1.616$}. Association of modes with particular irreps is performed by comparing the traces of submatrices within matrices $\M{D}$  (in this case simply the values on the diagonal of matrices $\M{D}$) and values in character table. This association is highlighted by colored squares. Here \mbox{$\left\{ \alpha \right\} = \left\{ \T{B}_2, \T{A}_1, \T{A}_2, \T{B}_1, \T{B}_2\right\}$}.}
	\label{fig:simpleStructure:I-D}
\end{figure}

The next step is the evaluation of matrices~$\M{D}(R)$ using~\eqref{eq:Dmatrix}. To do so, the impedance matrix~$\M{Z}$ is computed and \acp{CM} are found by~\eqref{eq:CMmat}. The resulting modal basis and corresponding matrices~$\M{D}(R)$ are depicted in Fig.~\ref{fig:simpleStructure:I-D} (the identity matrix for identity operation~$\T{E}$ is omitted). In this case, the diagonal blocks in~\eqref{eq:Dmatrix} are of size \mbox{$1 \times 1$} which is a signature of the abelian nature of the $\T{C}_{2\T{v}}$ symmetry group~\cite{McWeeny_GroupTheory}.

Collecting the unique traces from the on-diagonal block matrices (here directly from diagonal entries) gives the character table. Notice that the trace record corresponding to the $\T{B}_2$ irrep is repeated twice indicating that there are two modes in the matrix~$\Imat$ belonging to the $\T{B}_2$ irrep (the first and the fifth).

It is important to realize that diagonal blocks of the matrices~$\M{D}(R)$ and their corresponding traces are a property of given symmetry. Therefore, once calculated, they can be reused for all structures of the same point group.

%----------------------------------------------------------------------
\subsection{Construction of Symmetry-Adapted Basis~$\M{\Gamma}^\alpha$}
\label{sec:implementation:symAdaptBasis}

\begin{figure}
\includegraphics[width=\figwidth cm]{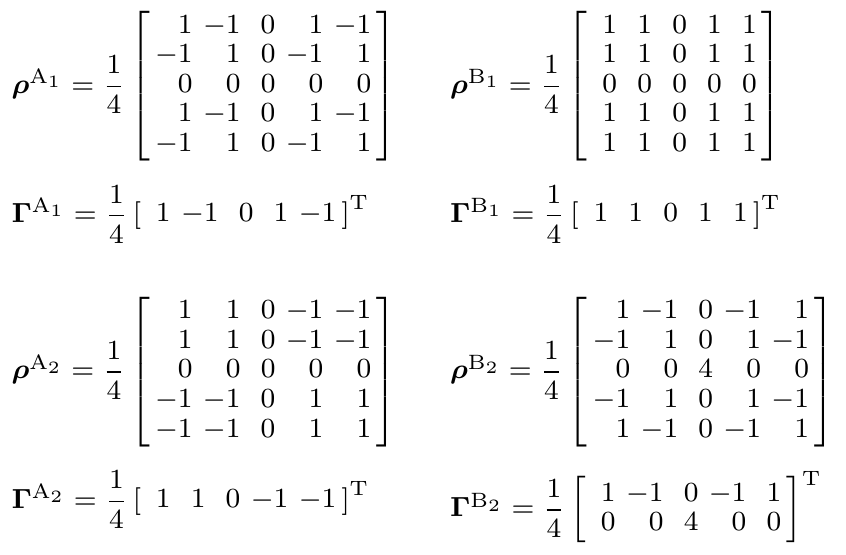}
\caption{Matrices~$\V{\rho}^\alpha$ and symmetry-adapted basis $\M{\Gamma}^\alpha$ for each irrep of the structure in  Fig.~\ref{fig:simpleStructure}.}
\label{fig:simpleStructure:rho-gamma}
\end{figure}

The final step is the construction of symmetry-adapted bases which are used for the block-diagonalization of the eigenvalue problem. The matrices~$\V{\rho}^\alpha_{i}$ are generated from~\eqref{eq:symmetryAdaptedBasis} and only 
\begin{equation}
	\label{eq:unknownsInIrreps}
	M = \sum_\alpha g^\alpha \eta^\alpha,
\end{equation}
linearly independent columns are kept in matrices $\M{\Gamma}^\alpha_{i}$, as depicted in Fig.~\ref{fig:simpleStructure:rho-gamma}. For a given discretization, the symmetry-adapted bases~$\M{\Gamma}^\alpha_{i}$ are utilizable either to compress an arbitrary eigenvalue problem as in~\eqref{eq:reduced:GEP} in order to find solutions within a given irrep only, or to establish a current solutions fulfilling the constraints imposed by symmetries of a given irrep, \ie{},~\mbox{$\Ivec^\alpha_i = \M{\Gamma}^\alpha_{i} \widehat{\Ivec}$}.

%xxxxxxxxxxxxxxxxxxxxxxxxxxxxxxxxxxxxxxxxxxxxxxxxxxxxxxxxxxxxxxxxxxxxxx
\section{Examples}
\label{sec:examples}

Using the implementation techniques described in Section~\ref{sec:implementation}, here we give practical examples of the mode tracking framework developed in Section~\ref{sec:background}. The methodology is presented on examples of increasing complexity, namely:
\begin{enumerate}
	\item four dipoles (simple abelian, Section~\ref{sec:examples:fourDipoles}),
	\item an equilateral triangle (the simplest non-abelian,  Section~\ref{sec:examples:equilateralTriangle}),
	\item a cubic array (complex non-abelian, Section~\ref{sec:examples:bowtieCube}),
    \item a radiator above a ground plane (Section~\ref{sec:examples:GNDantenna}).
\end{enumerate}

Throughout this section the eigenvalues have only been sorted according to their values. Potential vertical shifts in the modal spectrum, which are caused by an abrupt appearance or disappearance of a mode, have been treated according to Appendix~\ref{sec:app:jumpsDetection}.

%----------------------------------------------------------------------
\subsection{Example 1: Four Dipoles}
\label{sec:examples:fourDipoles}

Characteristic modes of a rectangular array of four dipoles treated in Fig.~\ref{fig:trackingDemonstration} are evaluated in this section. The geometry and meshing of the array are described in Section~\ref{sec:background:charModes}. The point symmetry group of this object is identical to that of Fig.~\ref{fig:simpleStructure}, \ie{}, \mbox{$G = \left\lbrace \T{E}, \sigma_\T{v}^{yz}, \sigma_\T{v}^{xz}, \T{C}_{2}^z\right\rbrace$}. The point group is abelian with four irreps, see character table in Fig.~\ref{fig:simpleStructure:I-D}, therefore, only non-degenerate modes exist in each irrep. From the point of view of Section~\ref{sec:background:symAdaptBasis}, the symmetry-adapted eigenvalue problems~\eqref{eq:reduced:GEP} give non-degenerated modes.

\begin{figure}[t]
    \includegraphics[width=\figwidth cm]{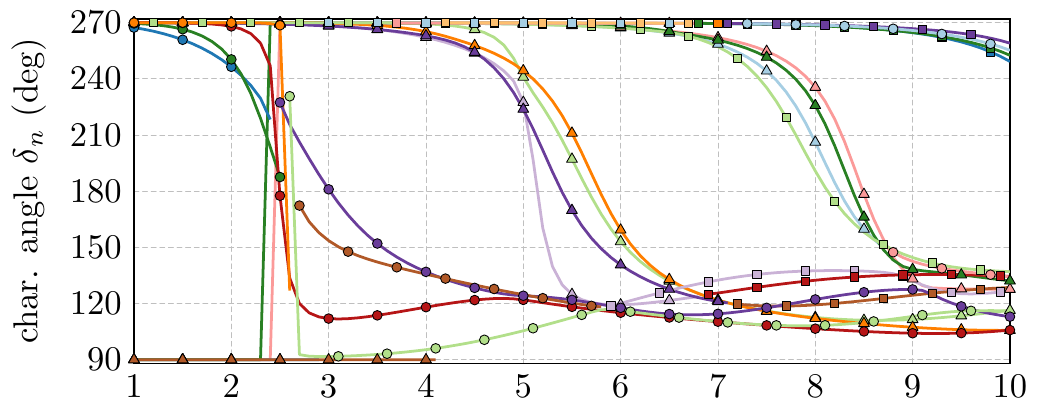}\\
    \animategraphics[loop, controls, autoplay, buttonsize=0.75em,width=\figwidth cm]{1}{Fig05b_fourDipoles_eigenAngle-animation}{0}{4}%
    \caption{An animation (available in Adobe Acrobat Reader) of tracked characteristic angles of the four-dipole array treated in Fig.~\ref{fig:trackingDemonstration}.
    Four modes of~\eqref{eq:reduced:GEP} with the smallest eigenvalue have been computed for every irrep of the corresponding symmetry point group. Collection of all these modes and their tracking via correlation-based algorithm is shown in the top panel. The bottom panel shows sorted characteristic angles when each irrep was solved separately via procedure described in this paper. Markers on every fifth frequency sample are depicted for clarity.}
    \label{fig:fourDipoles:eigenAngle}
\end{figure}

Tracked modes for this structure are shown in Fig.~\ref{fig:fourDipoles:eigenAngle}. The top figure shows results from correlation-based tracking while in the bottom figure, each color corresponds to one irrep (to a given~$\alpha$ in~\eqref{eq:reduced:GEP}) and modes within the irrep were tracked using the symmetry-based algorithm proposed in this paper. Note that the correlation-based algorithm has problem with proper connections of traces of characteristic angles near $ka \approx 2.6$ and that some modes are connected incorrectly (red, orange and light green lines in top figure). These issues are not present in the symmetry-based tracking. It can be observed that tracks are crossed only between modes from different irreps, \ie{}, modal crossings and modal crossing avoidances are no longer an issue. It is also important to stress that this simple procedure is less computationally expensive than correlation based tracking, see Section~\ref{sec:discussion:speedup}, and yields correct results irrespective of the spacing parameter~$d$. On the contrary, the correlation algorithm performs poorly for decreasing~$d$ and/or increasing number of dipoles.

Examples of eigencurrents of characteristic modes from each irrep are depicted in Table~\ref{tab:fourDipoles:eigenCurrent}.

\begin{table}[t] 
    \centering 
    \caption{Current distribution of the first two modes of each irrep of the four-dipole array. Only directions of currents on each dipole is depicted, different amplitudes are not considered.}
    
    \tikzstyle{myArrow} = [thick, line join = round, -stealth]
	\newcommand{\aH}{0.5cm} % arrow height
	\newcommand{\dX}{\hspace{0.2cm}}
	\newcommand{\dY}{0.02cm}
	\newcommand{\arrowU}{\tikz{\draw[myArrow] (0, -0.5*\aH) -> (0,  0.5*\aH);}\dX}
	\newcommand{\arrowD}{\tikz{\draw[myArrow] (0,  0.5*\aH) -> (0, -0.5*\aH);}\dX}
	\newcommand{\arrowI}{\tikz{\draw[myArrow] (0,  0.5*\aH) -> (0, \dY); \draw[myArrow] (0, -0.5*\aH) -> (0, -\dY);}\dX}
	\newcommand{\arrowO}{\tikz{\draw[myArrow] (0, \dY) -> (0, 0.5*\aH);  \draw[myArrow] (0, -\dY) -> (0, -0.5*\aH);}\dX}
	
    \begin{tabular}{
    	>{\centering\arraybackslash}m{0.6cm} >{\centering\arraybackslash}m{1cm}
        >{\centering\arraybackslash}m{1.2cm} >{\centering\arraybackslash}m{1cm}
        >{\centering\arraybackslash}m{1.2cm} >{\centering\arraybackslash}m{0.6cm}}
        irrep & $ka$ & $\M{I}_1$ & $\lambda_1$ & $\M{I}_2$ & $\lambda_2$\\ \toprule
        $\T{A}_2$ &2.8 & \arrowD \arrowD \arrowU \arrowU & $ 0.293$ & \arrowU \arrowD \arrowU \arrowD & $30.9$\\ 
        $\T{B}_2$ &2.8 & \arrowU \arrowU \arrowU \arrowU & $-0.313$ & \arrowU \arrowD \arrowD \arrowU & $2.15$\\
        $\T{A}_1$ &5.6 & \arrowO \arrowO \arrowO \arrowO & $-0.375$ & \arrowI \arrowO \arrowO \arrowI & $0.332$\\
        $\T{B}_1$ &5.6 & \arrowO \arrowO \arrowI \arrowI & $-0.119$ & \arrowO \arrowI \arrowO \arrowI & $1.60$\\ \bottomrule
    \end{tabular}
    \label{tab:fourDipoles:eigenCurrent}
\end{table}

%----------------------------------------------------------------------
\subsection{Example 2: Equilateral Triangle}
\label{sec:examples:equilateralTriangle}

The equilateral triangle with side length~$\ell$ was chosen as the simplest example of a non-abelian symmetric structure. Equilateral triangles have the point symmetry group $\T{C}_{3\T{v}}$ which possesses two one-dimensional and one two-dimensional irreps, see Table~\ref{tab:charTableC3v} for its character table. According to group theory~\cite{McWeeny_GroupTheory}, the modes belonging to this two-dimensional irrep $\T{E}$ exist in degenerated pairs with identical eigenvalues at all frequencies. Thus, there are two eigenvalue problems~\eqref{eq:reduced:GEP} for $\alpha = \T{E}$, each of them giving one member of the completely degenerated modal pair. It is worth noting that the separation into the degenerated pair (the form of matrix $\V{\Gamma}_i^\alpha$) is independent of frequency, \ie{}, modal tracking is possible even within this pair.

\begin{figure}[t]
	\centering
    \subfloat[irrep $\T{A}_1$ ($\lambda = -922.9$)]{%
       \includegraphics[width = \subfigwidth cm]{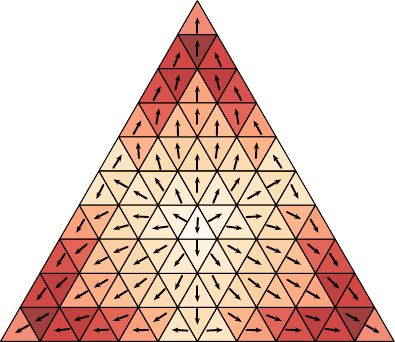}}
    \hfill
  	\subfloat[irrep $\T{A}_2$ ($\lambda = 26.49$)]{%
       \includegraphics[width = \subfigwidth cm]{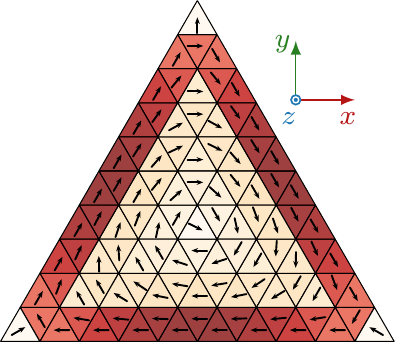}}
    \\
  	\subfloat[irrep $\T{E}$ ($\lambda = -4.893$)]{%
       \includegraphics[width = \subfigwidth cm]{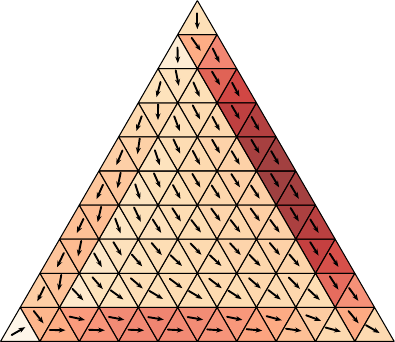}\quad
       \includegraphics[width = \subfigwidth cm]{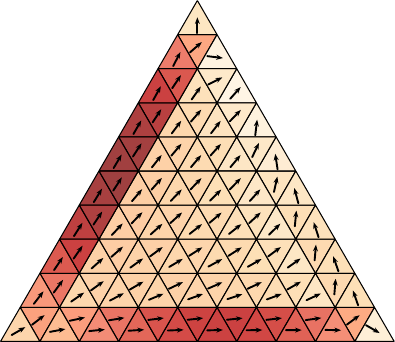}}
	\caption{First characteristic modes of each irreducible representation of equilateral triangle at electrical size $ka = 1$.}
	\label{fig:equilateralTriangle:eigenCurrent}
\end{figure}

Despite the aforementioned degeneracy, the treatment of modal crossings can be performed in exactly the same manner as for abelian point groups. The result of this procedure is depicted in Fig.~\ref{fig:equilateralTriangle:eigenAngle} where the lines belonging to irrep $\T{E}$ are twice degenerated. Examples of eigencurrents of characteristic modes from each irrep are depicted in Fig.~\ref{fig:equilateralTriangle:eigenCurrent}.

\begin{figure}[t]
     \animategraphics[loop, controls, autoplay, buttonsize=0.75em,width=\figwidth cm]{1}{Fig06_equilateralTriangle_eigenAngle-animation}{0}{3}
     \caption{An animation (available in Adobe Acrobat Reader) of six tracked characteristic angles on an equilateral triangle with the lowest eigenvalue. Each color represents one irrep. Red lines with square markers, which correspond to irrep~$\T{E}$, twice degenerated.}
	\label{fig:equilateralTriangle:eigenAngle}
\end{figure}

%----------------------------------------------------------------------
\subsection{Example 3: A Cubic Array}
\label{sec:examples:bowtieCube}

\begin{figure}[t]
	\centering
	\includegraphics[]{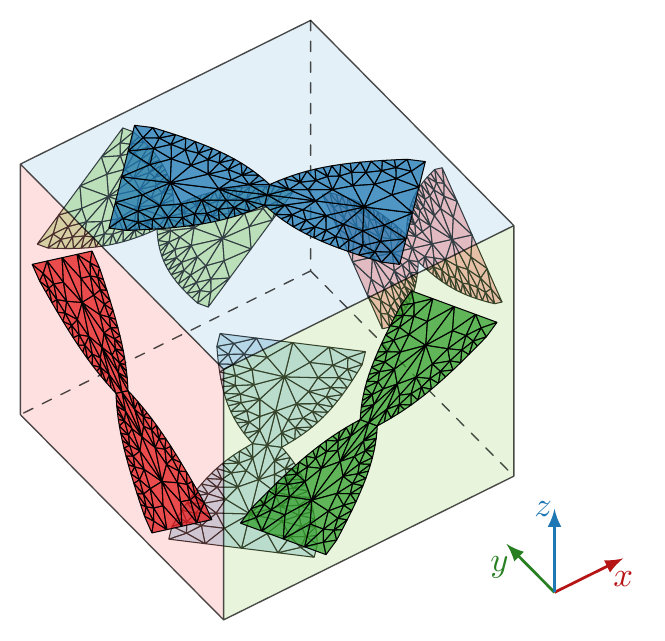}
    \caption{Six bowtie antennas arranged to a cubic array belonging to the $\T{T}_\T{d}$ point symmetry group.}
    \label{fig:bowtieCube:mesh}
\end{figure}

As a complex example, we examine the 3D cubic array of~bowtie antennas depicted in Fig.~\ref{fig:bowtieCube:mesh}. The point symmetry group of this object is~$\T{T}_\T{d}$ with character table shown in Table~\ref{tab:charTableTd} in Appendix~\ref{sec:app:characterTables}. Modes of this structure exhibit three-dimensional irreps and an intricate modal spectrum which is shown in Fig.~\ref{fig:bowtieCube:eigenAngle} and Fig.~\ref{fig:bowtieCube:eigenAngle:zoom}.

The complexity of the modal spectrum is enormous, with many modal crossings and crossing avoidances, see Fig.~\ref{fig:bowtieCube:eigenAngle}. Yet, the framework developed in this paper properly deals with all of them via a straightforward tracking procedure, see Figs.~\ref{fig:bowtieCube:eigenAngle} and~\ref{fig:bowtieCube:eigenAngle:zoom}.

\begin{figure}[t]
     \animategraphics[loop, controls, autoplay, buttonsize=0.75em,width=\figwidth cm]{1}{Fig09_bowtieCube_eigenAngle-animation}{0}{5}
     \caption{An animation (available in Adobe Acrobat Reader) of tracked characteristic angles of a cubic array depicted in Fig.~\ref{fig:bowtieCube:mesh}. Four modes with the lowest eigenvalue (not counting degeneracies) are shown for every irrep. Lines corresponding to irrep~$\T{E}$, are twice degenerated, lines corresponding to irreps $\T{T}_1$ and $\T{T}_2$ are three times degenerated. The highlighted area is enlarged in Fig.~\ref{fig:bowtieCube:eigenAngle:zoom}. Only every fifth marker is depicted for clarity.}
	\label{fig:bowtieCube:eigenAngle}
\end{figure}

\begin{figure}[t]
	\includegraphics[width=\figwidth cm]{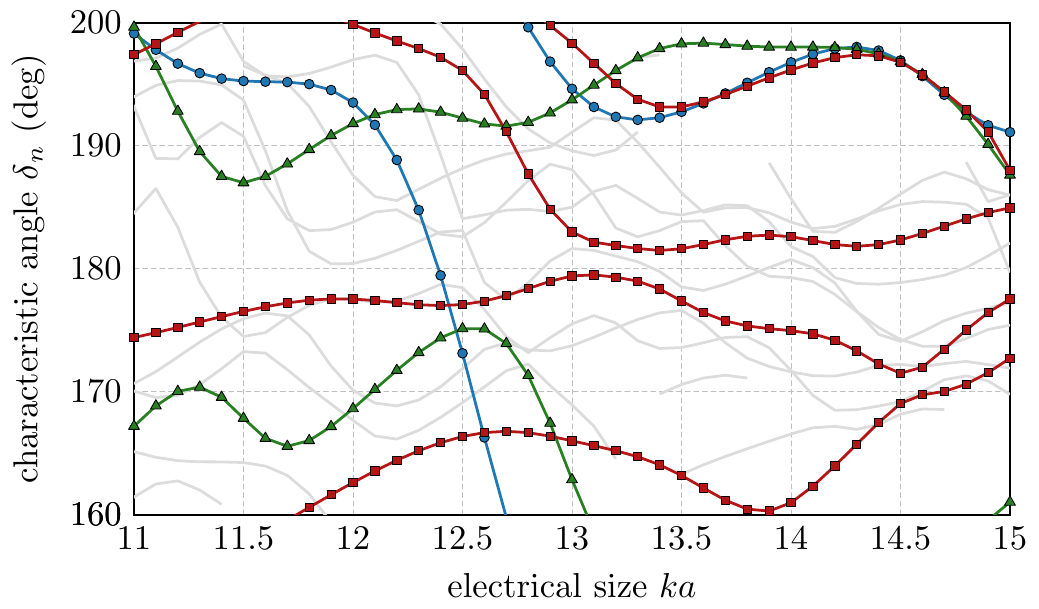}
	\caption{Enlarged area from Fig.~\ref{fig:bowtieCube:eigenAngle}. The traces depicted by the same color cannot cross as they belong to the same irrep. Modes of irreps $\T{T}_1$ and $\T{T}_2$ are plotted in gray color for clarity.}
	\label{fig:bowtieCube:eigenAngle:zoom}
\end{figure}

%----------------------------------------------------------------------
\subsection{Characteristic Modes of a Radiator Above a Ground Plane}
\label{sec:examples:GNDantenna}

\begin{figure}[t]
\centering 
	\includegraphics[]{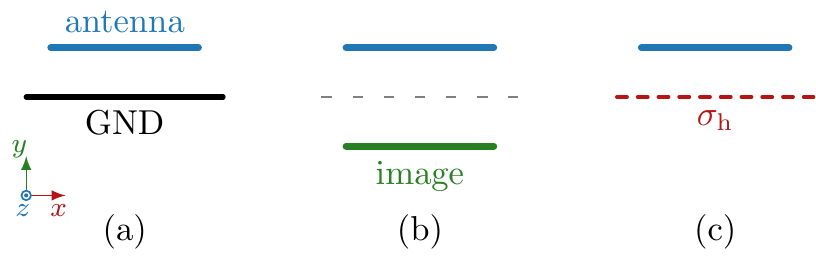} % width=\figwidth cm
	\caption{(a) Original arrangement of an antenna motif above ground plane GND, (b) removal of the ground plane via image theory, and (c) point group~$\T{C}_\T{s}$ with symmetry operator~$\sigma_\T{h}$ (for the character table, see Table~\ref{tab:GNDvsPhysSym}).}
	\label{fig:GNDreplacedBySym}
\end{figure}

The procedure presented in this paper can advantageously be applied to scenarios where an antenna is placed over a \ac{PEC} ground plane, see Fig.~\ref{fig:GNDreplacedBySym}a, and only the characteristic modes of the antenna are needed. This task is often approximated by replacing the ground plane with an image of the antenna~\cite{MartaEva_TheTCMRevisited}, see Fig.~\ref{fig:GNDreplacedBySym}b. In such a case, only the odd modes have to be extracted, since the even modes are not valid solutions to the original problem depicted in Fig.~\ref{fig:GNDreplacedBySym}a.

\begin{table}[t] 
    \centering 
    \caption{Character table of point group~$\T{C}_\T{s}$ (first three columns) \cite{McWeeny_GroupTheory} and characterization of modes and physical symmetries with respect to the modes belonging to irreps~$\T{A}'$ and $\T{A}''$.}
    \begin{tabular}{ccccc}
        $\T{C}_\T{s}$ & $\T{E}$ & $\sigma_\T{h}$ & solution & physical symmetry (GND) \\ \toprule
        $\T{A}'$ & $+1$ & $+1$ & even ($\uparrow\uparrow$) & PMC \\ 
        $\T{A}''$ & $+1$ & $-1$ & odd ($\uparrow\downarrow$) & PEC \\ \bottomrule
    \end{tabular} 
    \label{tab:GNDvsPhysSym}
\end{table}

The same task can be solved via point group theory. The arrangement in Fig.~\ref{fig:GNDreplacedBySym}a is, in fact, equivalent to the one depicted in Fig.~\ref{fig:GNDreplacedBySym}c, \ie{}, the presence of a physical symmetry can be replaced by the~$\T{C}_\T{s}$ symmetry group, see Table~\ref{tab:GNDvsPhysSym} for its character table. The symmetry operator~$\sigma_\T{h}$ splits the spectrum of modes into two irreps, $\T{A}'$ and $\T{A}''$. Irrep~ $\T{A}'$ contains only even modes, \ie{}, those corresponding to the presence of PMC, while irrep~$\T{A}''$ contains all the remaining (odd) modes, \ie{}, those corresponding to the presence of PEC.

To gather all antenna modes over the \ac{PEC} symmetry plane, the symmetry-adapted basis~$\M{\Gamma}^{\T{A''}}$ is constructed and the impedance matrix~$\M{Z}$ describing the original problem is reduced as
\begin{equation}
    \label{eq:GNDCMs_eq1repl}
    \widehat{\M{Z}}^{\T{A}''} = \left(\M{\Gamma}^{\T{A''}}\right)^\trans \M{Z} \M{\Gamma}^{\T{A''}}
\end{equation}
and substituted into~\eqref{eq:reduced:GEP} with a particular choice of~\mbox{$\widehat{\M{A}}_i^\alpha = \IM\{\widehat{\M{Z}}^{\T{A}''}\}$} and \mbox{$\widehat{\M{B}}_i^\alpha = \RE\{\widehat{\M{Z}}^{\T{A}''}\}$} respectively. A favorable aspect of this procedure is a remarkable speed-up as compared to the evaluation of the problem from Fig.~\ref{fig:GNDreplacedBySym}b. Notice that the same effect can be achieved via the incorporation of \ac{PEC} or \ac{PMC} boundary conditions directly into the Green's function of the problem~\cite{feko}. With this modified Green's function, the system matrix within the \ac{MoM} framework will attain exactly the same form as in~\eqref{eq:GNDCMs_eq1repl}.

%xxxxxxxxxxxxxxxxxxxxxxxxxxxxxxxxxxxxxxxxxxxxxxxxxxxxxxxxxxxxxxxxxxxxxx
\section{Discussion}
\label{sec:discussion}

Point group theory establishes a solid theoretical background for the thorough discussion of several questions related to modal decomposition and mode tracking in general. Some recurring questions in this area (yet-to-be fully answered) are discussed in the following section from the perspective of point group theory.

%----------------------------------------------------------------------
\subsection{On the Physical Importance of Modal Tracking}
\label{sec:discusion:importance}

This section addresses the question of whether modal tracking is or is not needed. Central to this discussion is the comparison of modal interpretations in the time and time-harmonic (frequency) domains.

\subsubsection{Tracking in the Time-Harmonic Domain}
\label{sec:discussion:timeDomain}

\begin{figure}[t]
	\includegraphics[width=\figwidth cm]{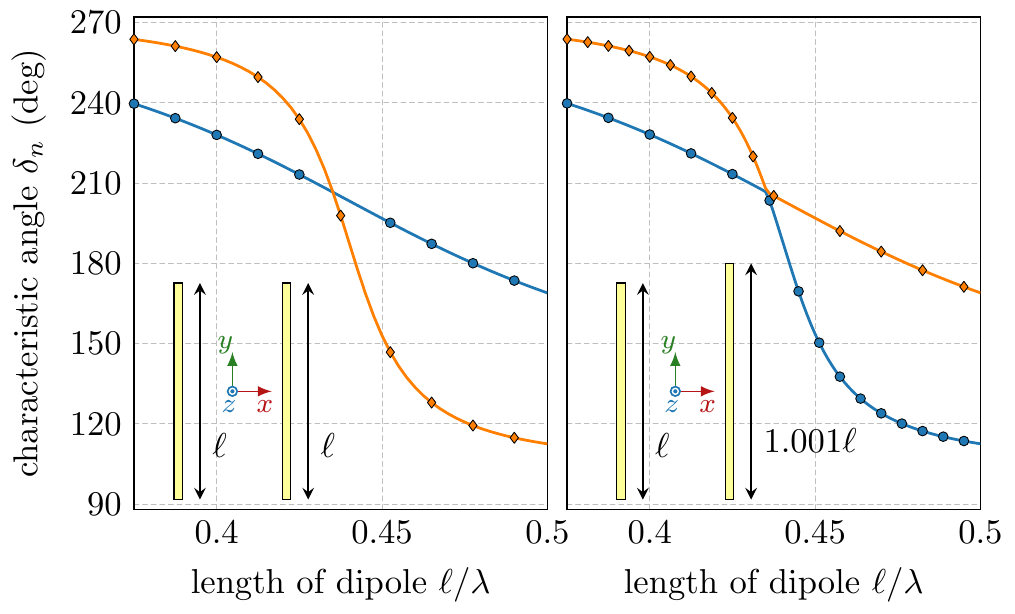}
	\caption{The first odd and even characteristic modes in the vicinity of their resonance. In the left panel, the dipoles have the same length, in the right panel the lengths are slightly different.}
	\label{fig:twoDipoles:eigenAngle}
\end{figure}

An interesting dilemma associated with modal tracking is depicted in Fig.~\ref{fig:twoDipoles:eigenAngle}. Both panels show very similar situations consisting of two dipoles placed in parallel. The dipoles in the left panel have exactly the same lengths~$\ell$. The dipoles in the right panel have lengths~$\ell$ and~\mbox{$\ell + \Delta\ell$}. Although the electromagnetic behavior of both structures must be very similar (especially as \mbox{$\Delta\ell \rightarrow 0$}), one can immediately notice qualitatively different results arising from symmetry-based mode tracking.  The dipoles with the same lengths are symmetrical, with the two plotted modal traces belonging to different irreps, and there we observe the expected eigenvalue trace crossing. The second setup (right panel) is not symmetrical, only the identity operation~$\T{E}$ exists (point group $\T{C}_1$), and therefore all characteristic modes belong to the same irrep so no modal traces can cross (as dictated by the von Neumann-Wigner theorem).

Owing to the arbitrarily small alteration which may affect this discrete change in behavior, we may be led to the conclusion that no physical phenomenon should depend on whether the eigenvalues as functions of frequency cross or not. It should be sufficient to interpret modal data independently at each frequency\footnote{This argument has already been raised by R.~Mittra during the IEEE AP Symposium in 2016.}. Such a conclusion is correct if modes at a single frequency are desired or if summation formulas such as~\cite{HarringtonMautz_TheoryOfCharacteristicModesForConductingBodies}
\begin{equation}
\label{eq:Disc:sum1}
\M{I}\left( \omega  \right) = \sum\limits_n a_n \left( \omega  \right) \M{I}_n \left( \omega  \right)
\end{equation}
are used. If so, mode ordering is irrelevant. A different situation arises when one wants to perform~\eqref{eq:Disc:sum1} in the time domain or to study the impulse response of a single mode, \ie{}, $\mathcal{F}^{-1} \left\{ a_n \left( \omega  \right) \M{I}_n \left( \omega  \right) \right\}$, where $\mathcal{F}^{-1}$ denotes an inverse Fourier's transform. In such a case, it must be stated that the solution presented in Fig.~\ref{fig:twoDipoles:eigenAngle} is the only correct solution\footnote{If improperly tracked, the spectrum is discontinuous which leads to unphysical time-domain artifacts with violation of causality being a notable example.} for both arrangements irrespective of how close in appearance (how small $\Delta\ell$ is) the structures are to each other.

\subsubsection{Modal Currents in the Time Domain}
\begin{figure}[t]
	\includegraphics[width=\figwidth cm]{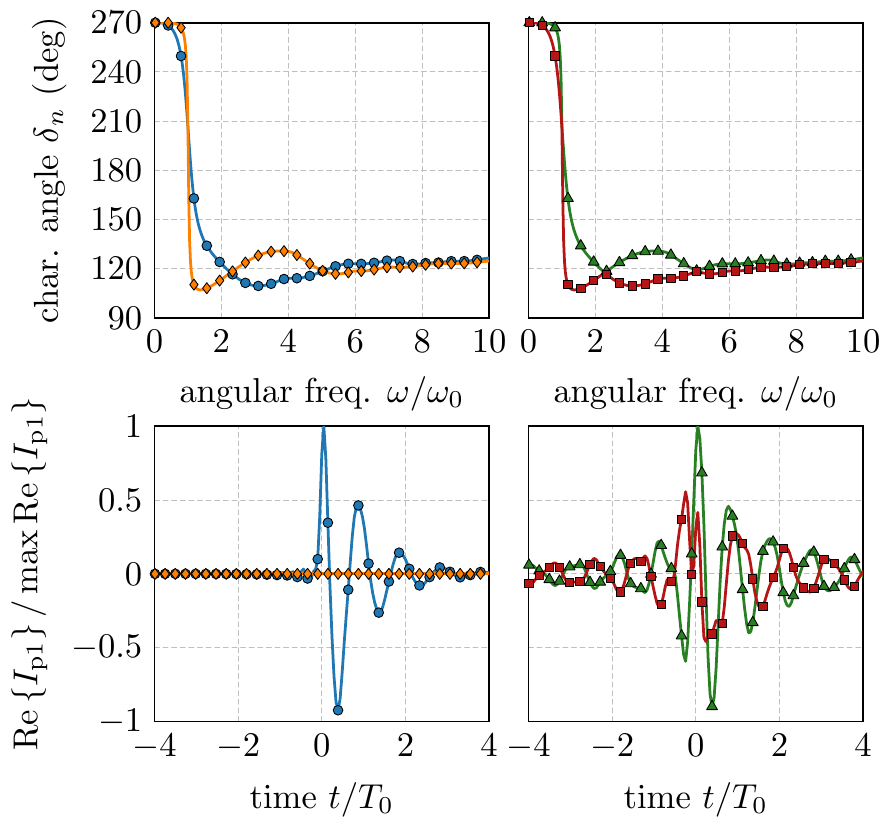}
	\caption{The influence of improper modal tracking on time domain characteristics. The left column represents modal tracking according to this paper. The right column represents improper tracking. Time domain responses show modal contributions of the two modes to the total current flowing through delta gaps feeding the structure.}
	\label{fig:twoDipoles:FFT}
\end{figure}

This subsection shows a simple example of the influence of improper modal tracking on time domain modal characteristics. As an example, two dipoles from the left panel of Fig.~\ref{fig:twoDipoles:eigenAngle} are used. The two most significant modes at first resonance were computed for electrical sizes ranging from \mbox{$ka = 0.04$} to \mbox{$ka = 16$} in $F = 400$ frequency samples. Dipoles were afterwards fed in their centers by two delta gap sources of the same orientation and amplitude across the entire frequency sweep. Modal contributions to the total time domain current at the source have been evaluated with an FFT algorithm after interpolation to $2^{17}$ frequency points of a double-sided spectrum and using a Blackman-Nuttall Window~\cite{Nuttall_SomeWindowsWithVeryGoodSIdelobeBehavior}. Prior to the FFT evaluation, eigencurrents were aligned at subsequent frequency samples. This has to be made because the direction of the eigencurrent is arbitrary with respect to~\eqref{eq:CMmat}. The results for properly tracked and improperly tracked modes are depicted in Fig.~\ref{fig:twoDipoles:FFT}. As previously argued, these data demonstrate how proper tracking ensures meaningful time domain responses of the individual modes.

%----------------------------------------------------------------------
\subsection{Effective Generation of Symmetry-Adapted Bases}
\label{sec:discussion:symAdaptBasis}

It is advantageous to use the \textit{a priori} tracking described in Section~\ref{sec:background:symAdaptBasis} and Section~\ref{sec:examples} owing to its straightforward implementation and capability of reducing the computational burden of the eigenvalue decomposition as shown by~\eqref{eq:eigCompTimeReduction}. However, to accomplish this, the symmetry-adapted bases~$\V{\Gamma}_i^\alpha$ must first be constructed. Consequently, one \Quot{pilot} eigenvalue decomposition~\eqref{eq:arbGEP} is needed to evaluate the $\DR$~matrices. One straightforward option is to take \mbox{$\M{A} \equiv \M{X}$} and \mbox{$\M{B} \equiv \V{\Psi}$}, where~\cite{JelinekCapek_OptimalCurrentsOnArbitrarilyShapedSurfaces}
\begin{equation}
	\V{\Psi} = \left[ \left\langle \basisFcn_m \left(\V{r}\right), \basisFcn_n \left(\V{r}\right) \right\rangle \right].
\end{equation}
As compared to the immediate choice of characteristic mode basis ($\M{X}$, $\M{R}$), the modes generated by ($\M{X}$, $\V{\Psi}$) have better numerical dynamics.

%----------------------------------------------------------------------
\subsection{Reduction of Computational Time With \textit{A Priori} Tracking}
\label{sec:discussion:speedup}

Utilization of \textit{a priori} tracking makes it possible to significantly accelerate modal decomposition. Assuming~\mbox{$M \gg 1$} discretization elements, the speed-up, expressed as a ratio between computational time~$T_\T{pg}$ needed to evaluate all reduced problems~\eqref{eq:reduced:GEP} and computational time~$T$ needed to evaluate the original problem~\eqref{eq:CMmat}, is expressed as
\begin{equation}
	\label{eq:eigCompTimeReduction}
	\frac{T_\T{pg}}{T} \propto \sum\limits_\alpha \frac{g^\alpha N^\alpha}{N} \left(\frac{\eta^\alpha}{M}\right)^2,
\end{equation}
where $N^\alpha$ is the number of modes in one degeneration of irrep~$\alpha$, \ie{}, \mbox{$N = \sum_\alpha g^\alpha N^\alpha$}. The relation between~$\eta^\alpha$ and~$M$ is mentioned in~\eqref{eq:unknownsInIrreps}. Note that \mbox{$g^\alpha N^\alpha \leq N$}, \mbox{$\eta^\alpha \leq M$}. Typical speed-ups for symmetry groups treated in this paper are depicted in Table~\ref{tab:AprioriTrackingSpeedUp}.

\begin{table}[t] 
  \centering 
  \caption{Reduction of computational time of characteristic mode decomposition for the \textit{a priori} tracking scheme. It is assumed that generalized Schur decomposition (QZ algorithm) is applied~\cite{Saad_NumericalMethodsForLargeEigenvalueProblems}.}
  \begin{tabular}{cccc}
    $G$ & $T_\T{pg}/T$ & $G$ & $T_\T{pg}/T$ \\ \toprule
    $\T{C}_1$ & $1$ & $\T{C}_\T{2v}$ & $1/2^4$ \\ 
    $\T{C}_\T{s}$ & $1/2^3$ & $\T{T}_\T{d}$ & $0.013$ \\ \bottomrule
  \end{tabular} 
  \label{tab:AprioriTrackingSpeedUp}
\end{table}

%----------------------------------------------------------------------
\subsection{Relationship between Geometrical and Physical Symmetries}
\label{sec:discussion:PECPMC}

With respect to the theory developed in this paper and the example in Section~\ref{sec:examples:GNDantenna}, it is interesting to point out that the correspondence between geometrical mirror symmetries and physical symmetries generated by \ac{PEC} and \ac{PMC} planes is general. Specifically, in the case of a polar vector, such as electric current density $\Jr$, the introduction of a \ac{PMC} plane is identical to a reflection plane, while the introduction of a \ac{PEC} plane is identical to a mirror plane with odd parity (a reflection plus change of a sign). In the case of pseudovectors, such as magnetic current density $\V{M}$, the roles of the \ac{PEC} and \ac{PMC} planes are interchanged. This means that in the case of abelian reflection groups (groups in which all operations can be generated as combinations of reflections), such as $\T{C}_{2\T{v}}$, there exists a ``physical'' equivalent of every irrep which is generated by the \ac{PEC} and \ac{PMC} planes. 

Let us take the $\T{C}_{2\T{v}}$ point symmetry group with \mbox{$G = \left\lbrace \T{E}, \sigma_\T{v}^{yz}, \sigma_\T{v}^{xz}, \T{C}_{2}^z = \sigma_\T{v}^{xz} \sigma_\T{v}^{yz}  \right\rbrace$} as an example. Characters of this group, see~Fig.~\ref{fig:simpleStructure:I-D}, can be interpreted as follows. A character with value $+1$ means that the representative current density is invariant with respect to a given set of reflections, \ie{}, the mirror planes can be substituted by \acp{PMC}. A character with value $-1$ means that the sign of the representative current must be inverted if the current is to stay invariant, \ie{}, such mirror planes can be substituted by \acp{PEC}. Using the aforementioned rules, a character table from Fig.~\ref{fig:simpleStructure:I-D} can be converted into Table~\ref{tab:PECPMC_tableC2v}, from which reduced simulations and calculations may be derived. This physical representation produces exactly the same modes as those generated by reduced eigenvalue problems~\eqref{eq:reduced:GEP} of the corresponding irreps.

\begin{table}[t]
	\caption{A physical realization of irreps generated by the $\T{C}_{2\T{v}}$ point group.}
	\label{tab:PECPMC_tableC2v}
	\begin{center}
		\begin{tabular}{lcc}
			$\T{C}_{2\T{v}}$ & plane $yz$ & plane $xz$ \\ \toprule
			$\T{A}_1$ & PMC & PMC \\
			$\T{A}_2$ & PEC & PEC \\
            $\T{B}_1$ & PEC & PMC \\
			$\T{B}_2$ & PMC & PEC \\ \bottomrule
		\end{tabular}
	\end{center}
\end{table}

%xxxxxxxxxxxxxxxxxxxxxxxxxxxxxxxxxxxxxxxxxxxxxxxxxxxxxxxxxxxxxxxxxxxxxx
\section{Conclusion}
\label{sec:conclusion}
A general framework utilizing point group theory for an arbitrarily parametrized eigenvalue problem was presented. A process uniquely classifying modes into irreducible representation was described. This classification method is independent of mesh density and requires only that the selected mesh and basis functions accurately represent the symmetry of the underlying structure. Applying the von~Neumann-Wigner theorem on these irreducible representations conclusively decides when traces of eigenvalues can or cannot cross and, thus, the problem of crossing and crossing avoidance is solved. An important consequence is that eigenvalue traces of modes belonging to a non-symmetric object can never cross. An understanding and treatment of degenerated modes was also provided.

Two approaches, \textit{a priori} and \textit{a posteriori}, were shown and their performances were demonstrated on characteristic mode decomposition. Examples of varying complexity confirmed the validity and robustness of the proposed method. The \textit{a priori} method was shown to considerably reduce computational time with respect to the standard eigenvalue solution.

Future efforts may extend the presented method to include translational symmetries. The devised framework might be utilized for the design of non-correlated feeding networks and for handling degeneracies when seeking sets of optimal currents. Advanced topics, such as chirality and its role in modal formalism, might also be rigorously addressed.

The robust detection of symmetries for arbitrary objects would allow for a fully automated implementation of the proposed procedure.

\appendices

%xxxxxxxxxxxxxxxxxxxxxxxxxxxxxxxxxxxxxxxxxxxxxxxxxxxxxxxxxxxxxxxxxxxxxx
\section{Character Tables}
\label{sec:app:characterTables}

Character tables of the structures introduced in Section~\ref{sec:examples} are presented here. The character table for group $\T{C}_{2\T{v}}$ is depicted in Fig.~\ref{fig:simpleStructure:I-D}.

\begin{table}[h!]
	\caption{Character table for point group $\T{C}_{3\T{v}}$, \cite{McWeeny_GroupTheory}.}
	\label{tab:charTableC3v}
	\begin{center}
		\begin{tabular}{lccc}
			$\T{C}_{3\T{v}}$ & $\T{E}$ & $2\T{C}_3$ & $3\sigma_\T{v}$ \\ \toprule
			$\T{A}_1$ & $+1$ & $+1$ & $+1$ \\
			$\T{A}_2$ & $+1$ & $+1$ & $-1$ \\
			$\T{E}$   & $+2$ & $-1$ & $0$ \\ \bottomrule
		\end{tabular}
	\end{center}
\end{table}

\begin{table}[h!]
	\caption{Character table for point group $\T{T}_{\T{d}}$, \cite{McWeeny_GroupTheory}.}
	\label{tab:charTableTd}
	\begin{center}
		\begin{tabular}{lccccc}
			$\T{T}_{\T{d}}$ & $\T{E}$ & $8\T{C}_3$ & $3\T{C}_2$ & $6\T{S}_4$ & $6\sigma_\T{d}$ \\ \toprule
			$\T{A}_1$ & $+1$ & $+1$ & $+1$ & $+1$ & $+1$ \\
			$\T{A}_2$ & $+1$ & $+1$ & $+1$ & $-1$ & $-1$ \\
            $\T{E}$   & $+2$ & $-1$ & $+2$ & $ 0$ & $ 0$ \\
            $\T{T}_1$ & $+3$ & $ 0$ & $-1$ & $+1$ & $-1$ \\
            $\T{T}_2$ & $+3$ & $ 0$ & $-1$ & $-1$ & $+1$ \\ \bottomrule
		\end{tabular}
	\end{center}
\end{table}

%xxxxxxxxxxxxxxxxxxxxxxxxxxxxxxxxxxxxxxxxxxxxxxxxxxxxxxxxxxxxxxxxxxxxxx
%----------------------------------------------------------------------
\section{Correction of Vertical Shifts in Symmetry-Based Tracking }
\label{sec:app:jumpsDetection}

\begin{figure}[t]
	\includegraphics[width=\figwidth cm]{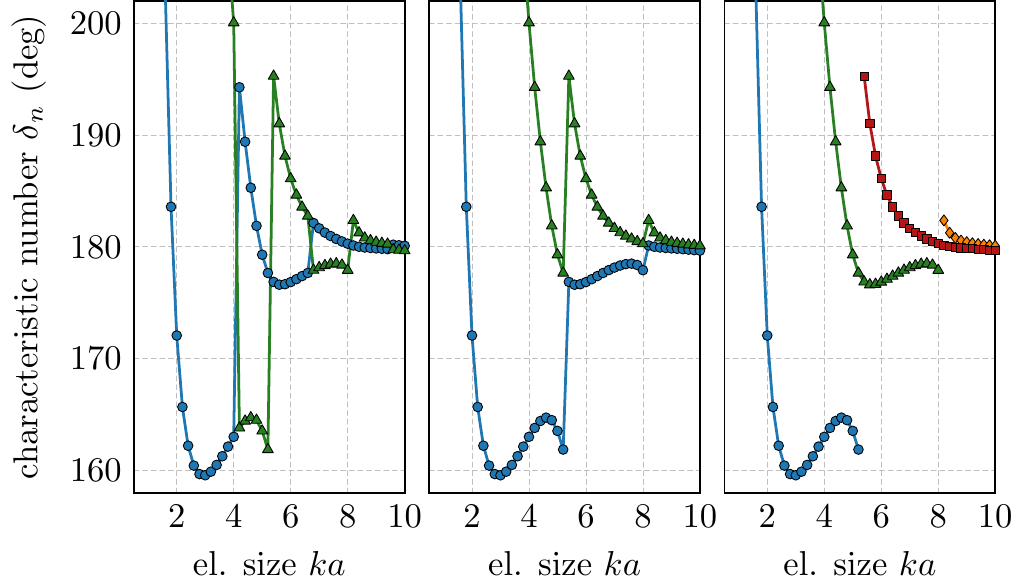}
	\caption{Two most significant modes at every frequency sample of irrep $\alpha = \T{E}$ of an equilateral triangle. The left panel shows raw unsorted data, the middle panel shows eigenvalues sorted with respect to their eigenvalues, and, finally the right panel shows data after the removal of vertical shifts.}
	\label{fig:equilateralTriangle_eigenAngle_jumpsTreatment}
\end{figure}

Under the assumption that all modes of the structure are available, the procedure described in this paper can track them flawlessly. In practice, however, we deal with a fixed (and usually small) number of the most significant (\eg{}, smallest eigenvalue magnitude) modes, \ie{}, $N \ll M$. Such a modal set naturally changes with frequency and, at specific frequencies, the appearance of a new mode and disappearance of an old mode occurs. When the tracking framework presented in this paper is used on this modal set, a~global vertical shift of modal tracks (in a~modal index) shows up every time a mode appears or disappears, see middle panel of Fig.~\ref{fig:equilateralTriangle_eigenAngle_jumpsTreatment}.

When frequency discretization is dense enough, the detection of the aforementioned shifts and their remedy can be based on the continuity of modal tracks, which is, in this paper, evaluated by 
\begin{equation}
	\kappa \left( \lambda^i_n \right) = \frac{{2{\lambda_n^i} - {\lambda_n^{i + 1}} - {\lambda_n^{i - 1}}}}{{2\max \left\{ {\left| {{\lambda_n^i}} \right|,\left| {{\lambda_n^{i + 1}}} \right|,\left| {{\lambda_n^{i - 1}}} \right|} \right\}}},
\end{equation}
where $i$ is a frequency index. A value of~\mbox{${\kappa} = 0.5$} is used as a detection threshold. After detecting a vertical shift, modal indices are globally shifted in order to minimize the value of~\mbox{$\kappa$}. The result of this procedure is shown in the right panel of Fig.~\ref{fig:equilateralTriangle_eigenAngle_jumpsTreatment} and throughout Section~\ref{sec:examples}.

% BIOGRAPHY
% =============================================================================
\begin{IEEEbiography}[{\includegraphics[width=1in,height=1.25in,clip,keepaspectratio]{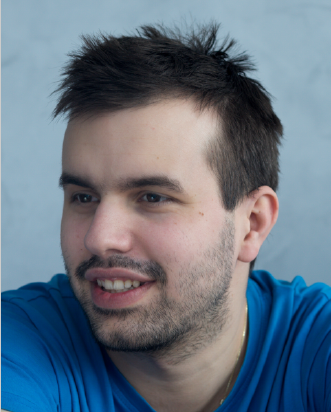}}]{Michal Masek}
received the M.Sc. degree in Electrical Engineering from Czech Technical University in Prague, Czech Republic, in 2015, where he is currently pursuing the Ph.D. degree in the area of modal tracking and characteristic modes. He is a member of the team developing the AToM (Antenna Toolbox for Matlab).
\end{IEEEbiography}

\begin{IEEEbiography}[{\includegraphics[width=1in,height=1.25in,clip,keepaspectratio]{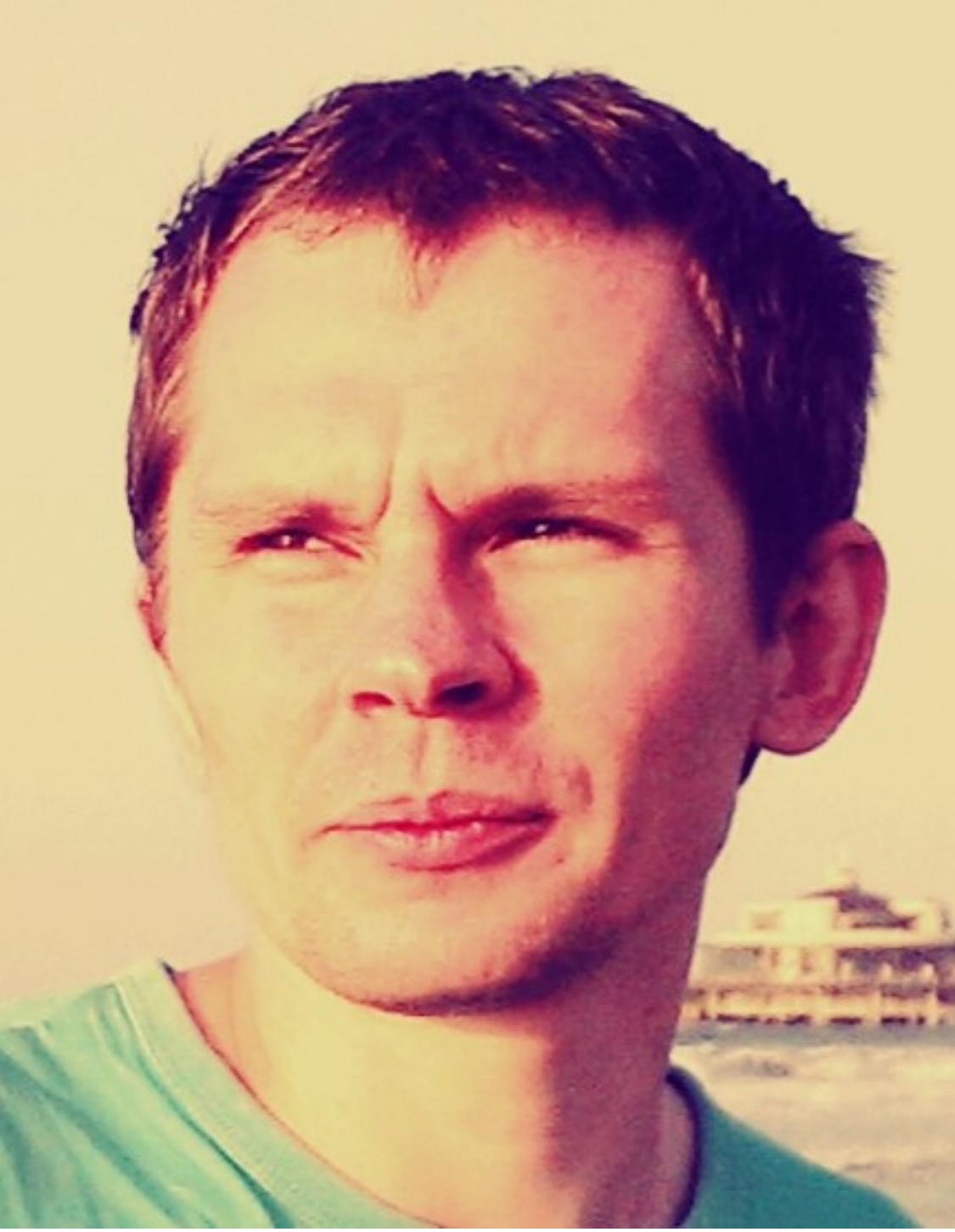}}]{Miloslav Capek}
(M'14, SM'17) received the M.Sc. degree in Electrical Engineering 2009, the Ph.D. degree in 2014, and was appointed Associate Professor in 2017, all from the Czech Technical University in Prague, Czech Republic.
	
He leads the development of the AToM (Antenna Toolbox for Matlab) package. His research interests are in the area of electromagnetic theory, electrically small antennas, numerical techniques, fractal geometry, and optimization. He authored or co-authored over 85 journal and conference papers.

Dr. Capek is member of Radioengineering Society, regional delegate of EurAAP, and Associate Editor of IET Microwaves, Antennas \& Propagation.
\end{IEEEbiography}

\begin{IEEEbiography}[{\includegraphics[width=1in,height=1.25in,clip,keepaspectratio]{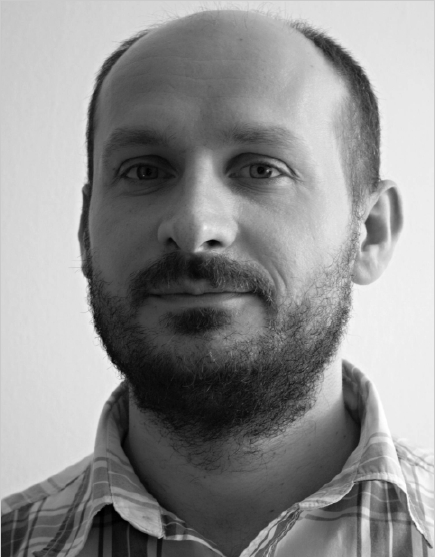}}]{Lukas Jelinek}
received his Ph.D. degree from the Czech Technical University in Prague, Czech Republic, in 2006. In 2015 he was appointed Associate Professor at the Department of Electromagnetic Field at the same university.

His research interests include wave propagation in complex media, general field theory, numerical techniques and optimization.
\end{IEEEbiography}

\begin{IEEEbiography}[{\includegraphics[width=1in,height=1.25in,clip,keepaspectratio]{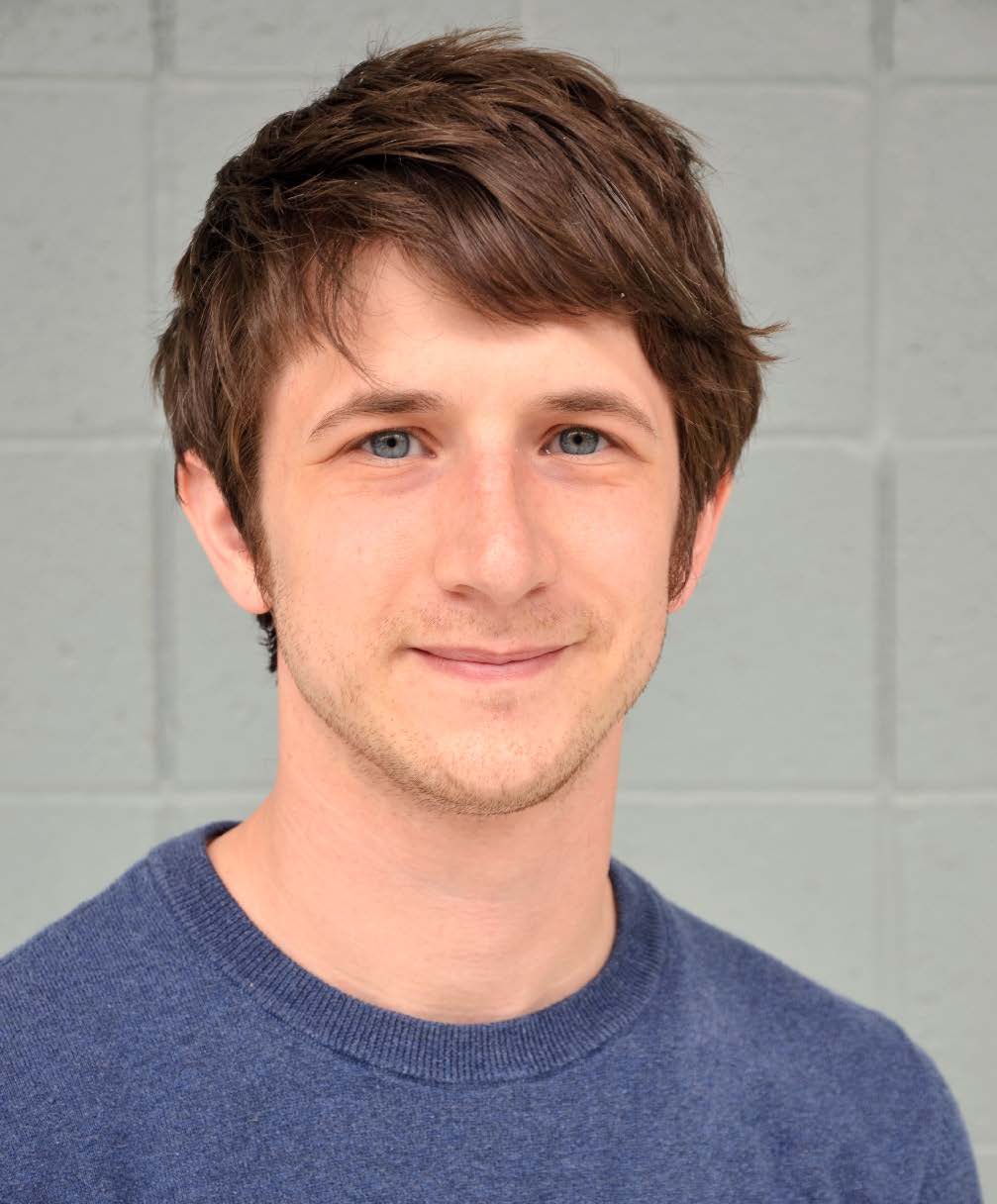}}]{Kurt Schab}(M'16)
Kurt Schab is an Assistant Professor of Electrical Engineering at Santa Clara University, Santa Clara, CA USA. He received the B.S. degree in electrical engineering and physics from Portland State University in 2011, and the M.S. and Ph.D. degrees in electrical engineering from the University of Illinois at Urbana-Champaign in 2013 and 2016, respectively.  From 2016 to 2018 he was an Intelligence Community Postdoctoral Research Scholar at North Carolina State University in Raleigh, North Carolina. His research focuses on the intersection of numerical methods, electromagnetic theory, and antenna design.  
\end{IEEEbiography}

\end{document}